\documentclass[]{spie}  

\usepackage{amsmath,amsfonts,amssymb}
\usepackage{graphicx}
\usepackage[colorlinks=true, allcolors=blue]{hyperref}

\title{The MICADO first light imager for the ELT: design and performance of the Focal Plane Mechanism prototype}

\author[a]{Kateryna Kravchenko}
\author[a]{Sebastian Rabien}
\author[a]{Matthias Deysenroth}
\author[a]{Luis Neumeier}
\author[a]{Lenard Spallek}
\author[a]{Mathias Honsberg}
\author[a]{Lothar Barl}
\author[a]{Julian Ziegleder}
\author[a]{Eckhard Sturm}
\author[a]{Richard Davies}
\affil[a]{Max Planck Institute for Extraterrestrial Physics, Giessenbachstr. 1, 85748 Garching bei M\"unchen, Germany}

\authorinfo{kkravchenko@mpe.mpg.de, phone: +49 89 30000 3949}

\begin{document} 
\maketitle

\begin{abstract}

MICADO is a first-generation instrument for the ELT. It will provide diffraction-limited imaging in standard, astrometric, and coronagraphic modes and long-slit spectroscopy at near-infrared wavelengths. The core of the MICADO instrument is its cryostat, which cools the internal optical and mechanical subsystems to 80 K. Following a light ray entering the cryostat through the entrance window, the first mechanism it encounters is the Focal Plane Mechanism. It consists of two independent movable devices mounted in one assembly: the aperture wheel and the focal plane wheel. The primary purpose of the aperture wheel is to rapidly block the light path, which is needed to mitigate persistence on the detectors. The focal plane wheel holds field stops, calibration masks, slits, and coronagraphs. The positioning requirements for the wheel are dominated by the coronographic masks demanding a 15~$\mu$m RMS repeatability. To fulfill this specification and avoid mechanical wear in the drive, a novel magnetically coupled gear system was developed at the Max Planck Institute for Extraterrestrial Physics (MPE). A magnetically coupled worm gear uses magnetic forces to transmit torque from the motor to the driven component without direct mechanical contact. This paper describes the design and performance of the magnetic drive and the first results of the focal plane wheel prototype tests in a cryogenic environment.

\end{abstract}


\section{INTRODUCTION}
\label{sec:intro}  

The Multi-AO Imaging Camera for Deep Observations \cite{MICADO} (MICADO) is a cryogenic near-infrared imager and spectrometer developed for the first-light operation at the 40-m Extremely Large Telescope \cite{10.1117/12.2631613} (ELT) of ESO. The instrument will be located on an ELT Nasmyth platform and utilize the optical straight-through port of the ELT's Pre-Focal Station \cite{2020SPIE11445E..1GL}. MICADO will operate with the MORFEO \cite{10.1117/12.2628969} multi-conjugate adaptive optics system MCAO as the primary adaptive optics mode and the single conjugate adaptive optics system SCAO \cite{SCAO}. The latter will be utilized during the early phases of instrument operations in the stand-alone configuration. MICADO will operate in wavelength range 0.8-2.4$\mu$m and will provide four observing modes: i) high- (HRI) and low-resolution (LRI) imaging with an un-vignetted field of view of 19“x19” at a pixel scale of 1.5~mas, and 50”x50” at a pixel scale of 4~mas, respectively; ii) astrometric imaging with astrometric precision of 50 $\mu$as relative to reference sources in the field; the long-slit spectroscopy with spectral resolution of R~$\sim$~20000; iv) high-contrast imaging enabled via a configuration of coronagraphs and Lyot stops. 

\begin{figure}[ht]
   \begin{center}
   \begin{tabular}{c} 
   \includegraphics[width=0.48\textwidth]{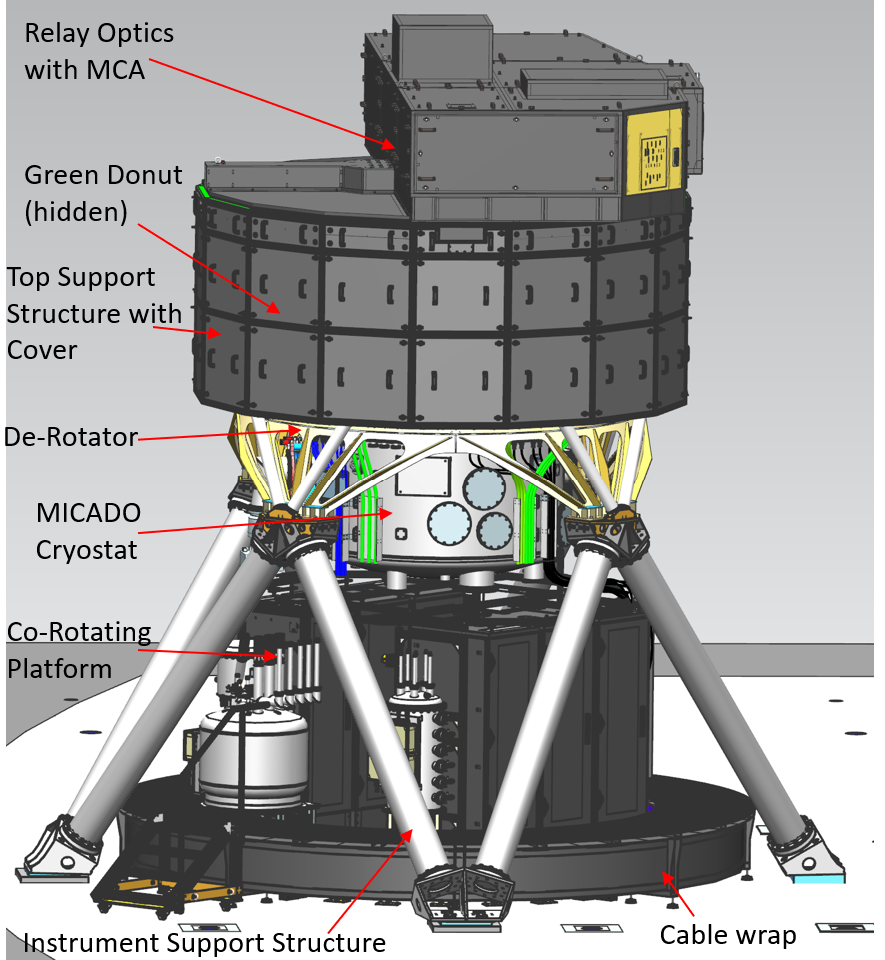}
   \includegraphics[width=0.49\textwidth]{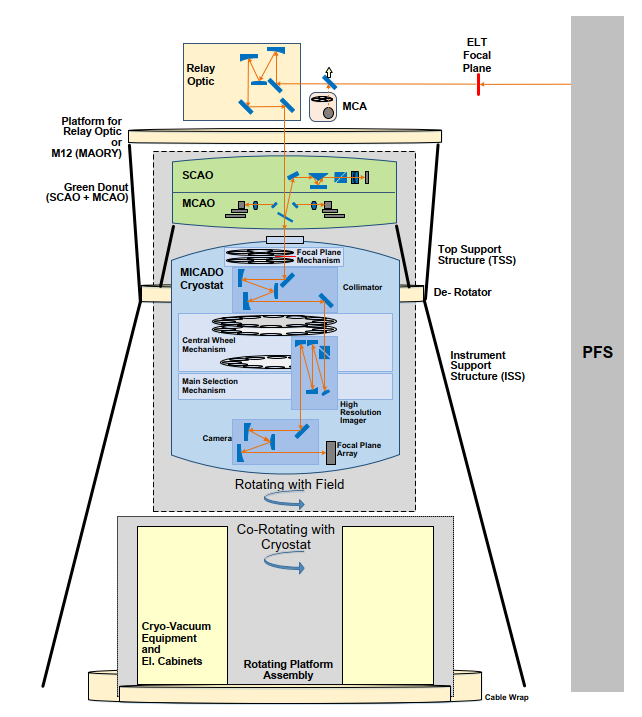}
   \end{tabular}
   \end{center}
   \caption{A three-dimensional view (left) and functional architecture (right) of MICADO in the stand-alone configuration. \label{fig:architecture}}
   \end{figure}

The MICADO architecture in its stand-alone configuration is shown in Fig.~\ref{fig:architecture}. The main building blocks of the instrument are the cryostat assembly~\cite{cryostat}, a de-rotator, two natural guide star wavefront sensor modules SCAO and MCAO, a relay optics~\cite{RO} and a calibration unit~\cite{MCA} (MCA). These sub-systems are integrated on a support structure over a co-rotating platform, which carries the electronics. A detailed description of the instrument architecture is provided in paper~13096-37 \cite{MICADO}. 

   \begin{figure}[h!]
   \begin{center}
   \begin{tabular}{c} 
   \includegraphics[width=0.6\textwidth]{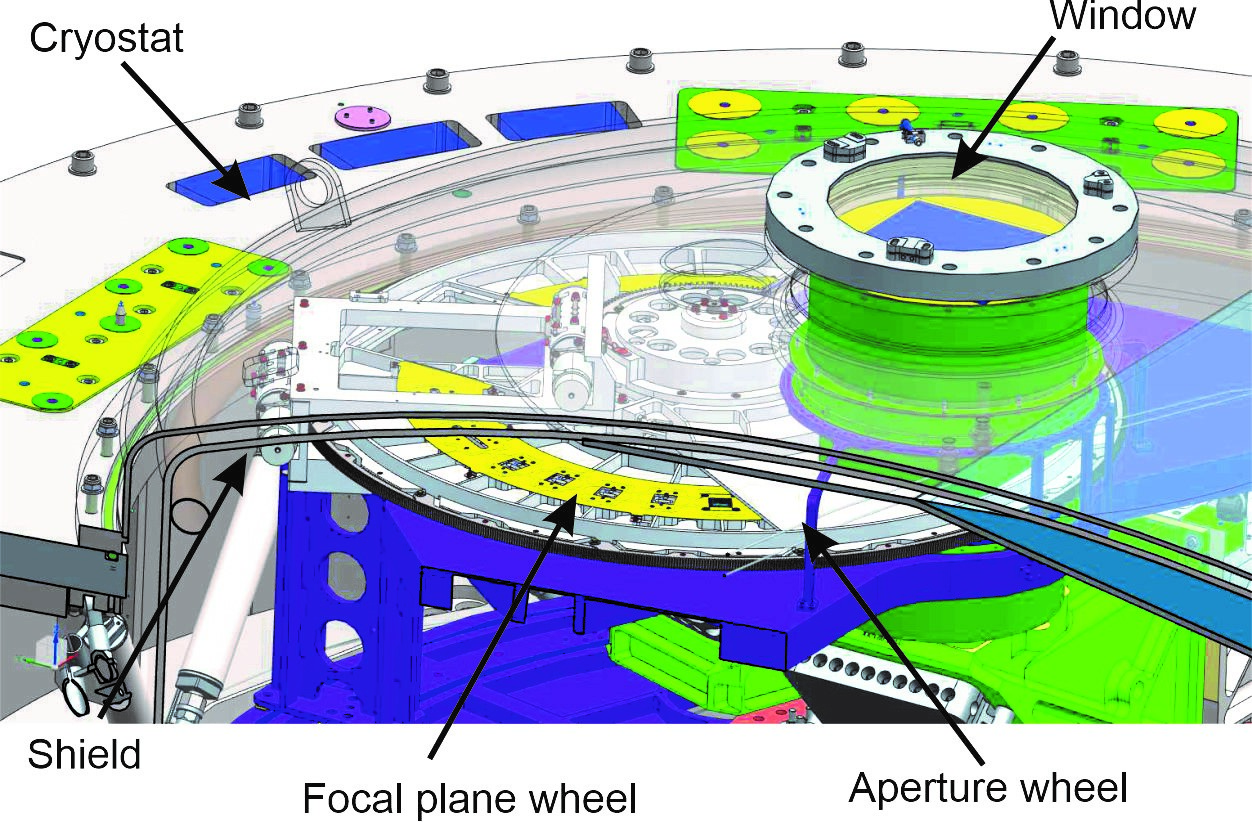}
   \end{tabular}
   \end{center}
   \caption{Location of the Focal Plane Mechanism inside the MICADO cryostat. \label{fig:cryostat_inside}}
   \end{figure}

The core of the MICADO instrument is the cryostat. This is a liquid nitrogen continuous flow cryostat which cools its internal subsystems to $\sim$80 K. It contains the instrument cold opto-mechanical sub-systems~\cite{10.1117/12.2311867}, mechanisms~\cite{MSM,2020SPIE11451E..3RR}, and detector array~\cite{detector,DPS}. The top view of the MICADO cryostat is shown in Figure~\ref{fig:cryostat_inside}. The first subsystem after the entrance window is the focal plane mechanism (FPM), where the relay optics re-images the ELT focal plane. The design and performance of the FPM prototype are the subject of this paper.

The overview of the FPM is provided in Sect.~\ref{FPM}. Sect.~\ref{MCWG} describes the magnetically coupled worm gear drive principle. The implementation of the drive in the FPM prototype and its cryogenic tests are described in Sect.~\ref{tests}.

   \begin{figure}[t!]
   \begin{center}
   \begin{tabular}{c} 
   \includegraphics[width=\textwidth]{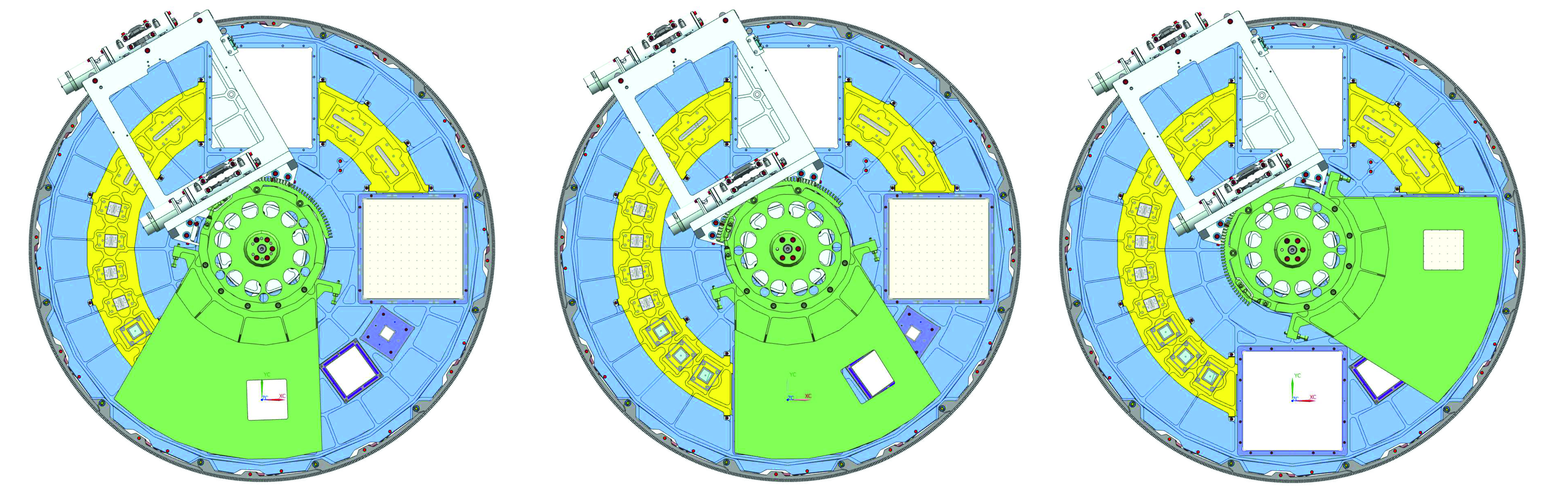}
   \end{tabular}
   \end{center}
   \caption{Overview of the Focal Plane Mechanism. The FPW holds various masks, apertures, coronographs, and slits (colored in blue). The AW is located on top of the FPW (colored in green). The AW offers three defined positions: a small field shadow (left panel), a closed position (middle panel) and the open position (right panel). \label{fig:aw}}
   \end{figure} 

   \begin{figure}[t!]
   \begin{center}
   \begin{tabular}{c} 
   \includegraphics[width=0.6\textwidth]{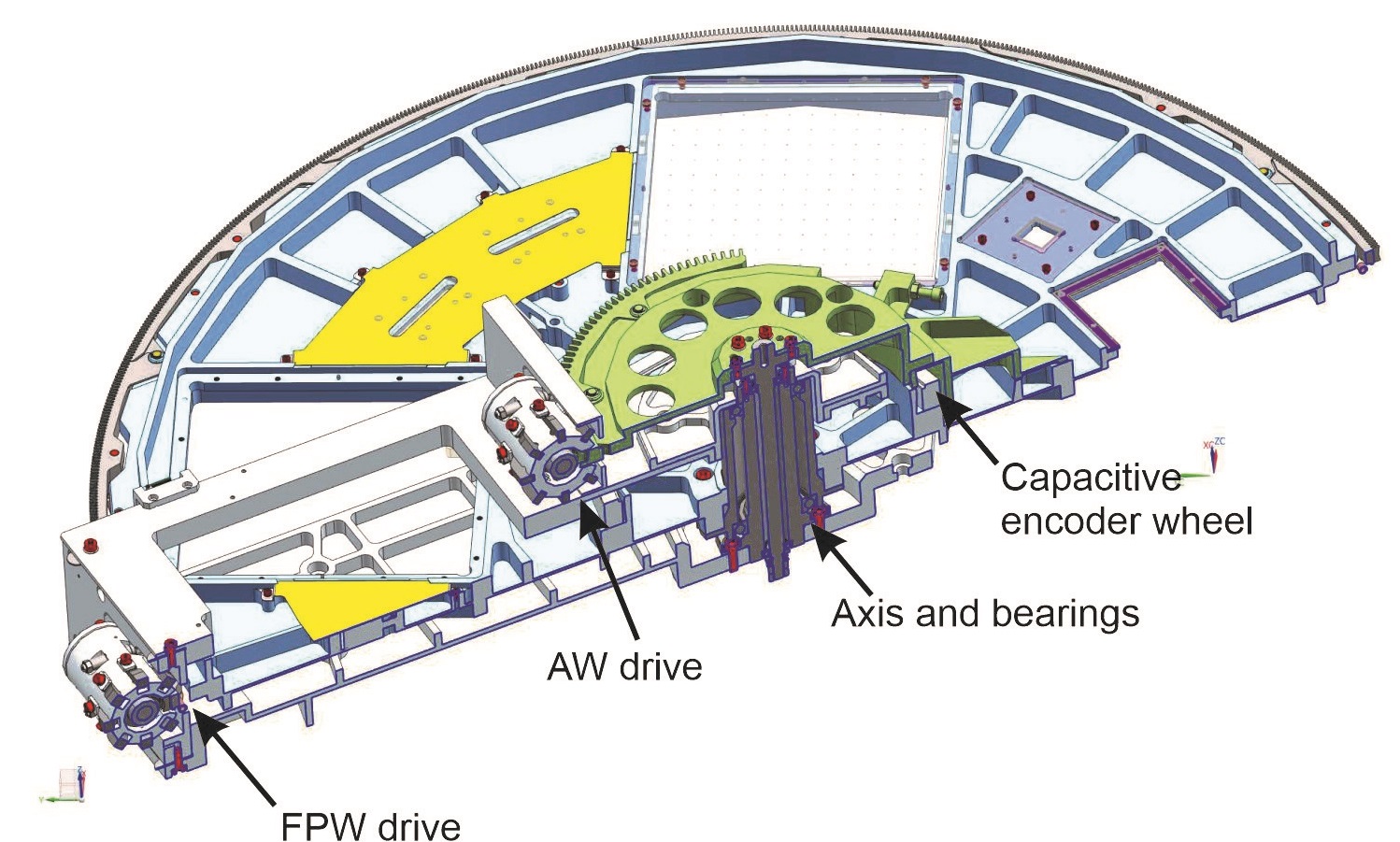}
   \end{tabular}
   \end{center}
   \caption{Detailed view of the FPW. \label{fig:fpw}}
   \end{figure} 

\section{The Focal Plane Mechanism}
\label{FPM}

The FPM is the first cryogenic mechanism inside the cryostat after the entrance window and the light baffle below it (Fig.~\ref{fig:cryostat_inside}). It comprises two parts mounted in one assembly: the aperture wheel (AW) and the focal plane wheel (FPW), as shown in Fig.~\ref{fig:aw}. Both wheels are supported by a central bearing system fixing the axis to the mounting structure. 

\subsection{Aperture wheel}

The AW is located above the cold optical train and can switch between three positions: closed, open, and a small aperture (Fig.~\ref{fig:aw}). This mechanism acts fast but at low precision. The requirement for the position repeatability of the AW is $\pm100\mu$m RMS. Its primary purpose is to block the light path (e.g., during presets or large offsets) as part of the persistence mitigation strategy. The small field aperture enables the optical elements on the FPW to be positioned close to each other. While the AW will offer three distinct positions for operation, the drive can move the wheel to any position thanks to the stepper motor. This allows fine-tuning the software-commanded three operational positions upon assembly and test. 
The shutter blade will be manufactured from aluminum~6061 or similar, matching the cold benches and the structural elements in the cryostat. The bearing system will consist of two spring-loaded shoulder ball bearings. The surfaces towards the entrance window will be left blank, and the surfaces towards the cold instrument will be blackened. 


   \begin{table}[ht]
\caption{List of the focal plane masks in MICADO.} 
\label{tab:masks}
\begin{center}       
\begin{tabular}{p{0.05\linewidth} | p{0.25\linewidth} | p{0.6\linewidth}}
\hline
\rule[-1ex]{0pt}{3.5ex}  Pos. & Mask & Short description  \\
\hline 
\rule[-1ex]{0pt}{3.5ex} 1	& Astrometric Pinhole mask	& Numerous holes across the large field to provide point sources for astrometric calibration of internal instrument distortions. \\
\rule[-1ex]{0pt}{3.5ex} 2	& High contrast field	& Transmits a field within the central detector for the HRI (6"x6"), pupil plane coronagraphy and SAM modes. \\
\rule[-1ex]{0pt}{3.5ex} 3	& Small field	& Transmits the full imaging field of the HRI (19"x19"). \\
\rule[-1ex]{0pt}{3.5ex} 4	& Large field	& Transmits the full imaging field of the LRI (50"x50"). \\
\rule[-1ex]{0pt}{3.5ex} 5	& CLC0	& Classical Lyot Coronagraph (CLC) optimized for small angular separation ($>$15 mas) in J, H band. \\
\rule[-1ex]{0pt}{3.5ex} 6	& CLC1	& CLC optimized for medium exoplanet separation ($>$30 mas) in J, H and K bands. \\
\rule[-1ex]{0pt}{3.5ex} 7	& CLC2	& CLC optimized for sensitivity at larger angular separation ($>$50 mas) for exoplanet and disk sciences in J, H and K bands. \\
\rule[-1ex]{0pt}{3.5ex} 8	& Wide slit	& Spectroscopic slit that is 3” long and $\sim$48mas wide, for IzJ and HK bands. \\
\rule[-1ex]{0pt}{3.5ex} 9	& Wide offset slit	& As for wide slit, position in focal plane offset for spectral dithering. \\
\rule[-1ex]{0pt}{3.5ex} 10	& Short slit	& Spectroscopic slit that is 3" long and 16mas wide, for IzJ band. \\
\rule[-1ex]{0pt}{3.5ex} 11	& Short offset slit	& As for short slit, position in focal plane offset for spectral dithering. \\
\rule[-1ex]{0pt}{3.5ex} 12	& Long slit	& Spectroscopic slit that is 15" long and 20mas wide, for J and HK bands. \\
\rule[-1ex]{0pt}{3.5ex} 13	& Long offset slit	& As for long slit, position in focal plane offset for spectral dithering. \\
\rule[-1ex]{0pt}{3.5ex} 14	& Ronchi mask / spare	& Technical mask, not used for operations. \\
\rule[-1ex]{0pt}{3.5ex} 15	& Pinhole slit	& Row of pinholes covering 15", for spectral calibration. \\
\rule[-1ex]{0pt}{3.5ex} 16	& Offset pinhole slit	& As for pinhole slit, but position shifted to match offset slits. \\
\hline 
\end{tabular}
\end{center}
\end{table}


\subsection{Focal plane wheel}

The FPW holds various focal plane masks, mainly small and large field masks for imaging, a pinhole mask for distortion calibration, three slits (each appearing twice to enable spectral dithering), and three coronagraphs. The full list of masks is listed in Table~\ref{tab:masks}. The FPW moves slower than the AW, but can be positioned very accurately to the desired imaging or spectroscopy mode. A detailed view of the FPW is shown in Fig.~\ref{fig:fpw}. A housing and baseplate hold the shafts of the bearing system comprised of a pair of spring-loaded shoulder ball bearings. 
The outer diameter of the wheel is roughly 894mm. The optical elements are placed at a radius of 290 mm, allowing sufficient space to allocate all focal plane masks of MICADO and foresee spare positions for additional components that may be needed in the future. Like the AW, the FPW will be manufactured from aluminum~6061 or similar.

Both the AW and the FPW will be equipped with two end switches and one reference switch realized via non-contact GMR sensors. The reference switch will be used to home the motor. Both wheels will be equipped with Phytron stepper motors with a resolver and driven in the closed loop. In the case of the FPW, a capacitive encoder system is foreseen to supply additional position information. This information will serve as an independent position measurement for the commissioning of the mechanism and can be provided as information for operation and testing. It is not foreseen to use the capacitive sensor signal for closed-loop operation.

Since the spectroscopy slits are positioned perpendicular to the radius of the FPW, the positioning requirement for the wheel is dominated by the coronographic masks, demanding a $\pm$15$\mu$m RMS repeatability. 
To fulfill this specification and avoid wear in the drive in the cryogenic environment, a novel magnetically coupled worm gear (MCWG) system was developed and tested at MPE. 

\section{Magnetically coupled worm gear}
\label{MCWG}

   \begin{figure}[t!]
   \begin{center}
   \begin{tabular}{c} 
   \includegraphics[width=0.5\textwidth]{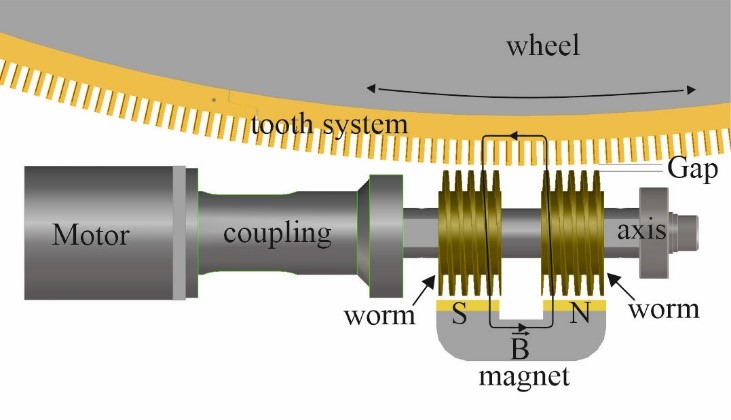}
   \includegraphics[width=0.45\textwidth]{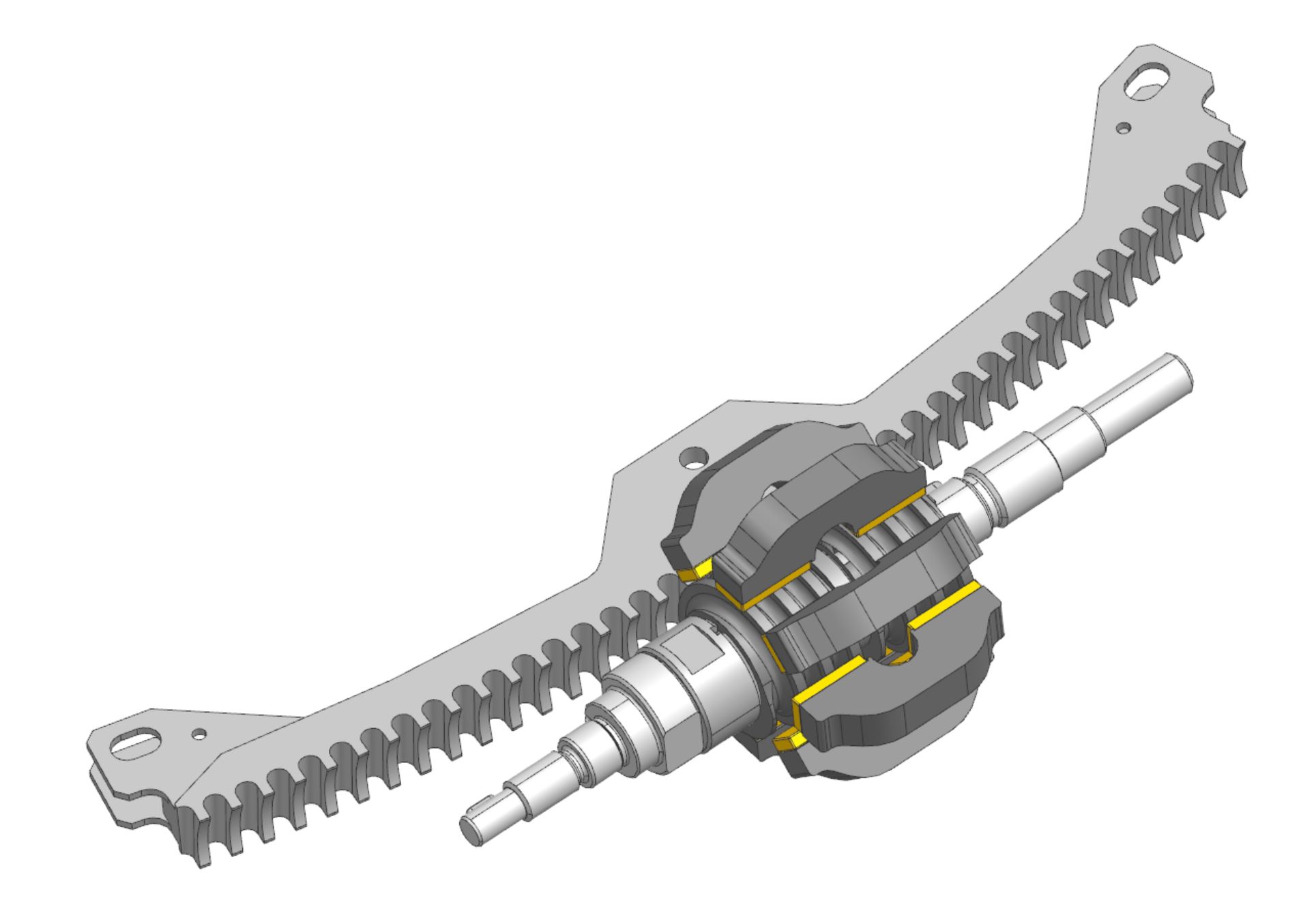}
   \end{tabular}
   \end{center}
   \caption{\textit{Left panel:} The working principle of the MCWG: magnetic forces transmit torque from the motor to the driven wheel without mechanical contact. \textit{Right panel:} Three-dimensional view of the MCWG design. \label{fig:mcwg}}
   \end{figure} 

Driving moving mechanisms in cryogenic environments with high accuracy is considerably more challenging than at room temperature. Achieving high-precision movement often requires a gear between the driving motor and the movable part. Nowadays, gears are often operated dry or with solid-state lubricant coatings. Utilizing mechanisms such as a rack and pinion drive in cryogenic conditions can result in wear, particle contamination, and limited lifetime. In order to avoid mechanical contact between the drive and the moving part, a magnetically coupled worm gear (MCWG) was developed at MPE. 

A magnetically coupled worm gear uses magnetic forces to transmit torque from the motor to the driven component without direct mechanical contact. The basic principle of the MCWG is sketched in Fig.~\ref{fig:mcwg}. The system consists of two main parts: the driving unit connected to the motor and the driven unit connected to the output shaft. The drive comprises two spiral-shaped worm elements mounted on a common non-magnetic axis rotated by the driving motor. 
The worm elements are enclosed by a permanent magnet system consisting of seven discrete U-shaped magnets, such that one worm always faces the south poles and the other - the north poles of the magnets. 
The driven wheel has a tooth system mounted to its edge with a pitch matching the worms' pitch. With the worms and the tooth system manufactured from a soft ferromagnetic material (ALLIEDPUREIRON 99.9\% iron), the magnet's field lines are closed over the tooth system. The size of the gap between the worms and the tooth system is kept small compared to the distance between individual teeth. In this arrangement, the tooth system (and thus the wheel) is dragged along with the rotation of the worm axis. The key advantage is that the torque is transmitted through magnetic forces rather than physical contact. This non-contact operation significantly reduces wear and particle generation, making the system ideal for vacuum and cryogenic environments.
 
The main properties of the MCWG drive are:
\begin{itemize}
    \item There is no contact between the worms and the wheel, so there are no wear, contamination, or lifetime issues;
    \item It is suited for clean room, cryogenic and space environments;
    \item It does not require any lubricant or special coating;
    \item The magnetic field is confined locally at the worm drive, so no freely moving magnets and open fields are present elsewhere;
    \item With relatively free choice of the pitch of the worm, a high gear ratio possible in a single stage;
    \item No motor current is required to hold a position resulting in zero heat load when not used;
    \item High resolution and repeatability of the positioning. The resolution of the mechanism is defined by the pitch of the worm and by the resolution of a stepper motor.
\end{itemize}

Being a novel technology, the MCWG has been filed for a patent application at the European Patent Office~\cite{mcwg}. It is implemented in both the AW and the FPW mechanisms. In the case of the AW, with a primary gear ratio of 50 and a 200 steps/rev motor, a single motor step moves the wheel by 0.036 degrees. In the case of the FPW configuration, the following parameters have been chosen: wheel diameter of 893.2~mm and tooth pitch of 6~mm, which converts into gear ratio of $\sim$470. A Phytron VSS 42.200 stepper motor (200 steps/revolution) with a resolver has been chosen for the drive. The mechanism's resolution is, then, 0.77 degrees for one full motor turn and 13.86 arcseconds for a single motor step. Those parameters convert into a 19.5$\mu$m/step at the optics radius of the wheel.

Due to the large diameter of the FPW and the coefficient of thermal expansion (CTE) mismatch between the soft magnetic iron teeth and the aluminium body, the tooth system mounted to the wheel is segmented. Each segment has a spring-loaded mount holding it to the aluminium wheel. Gaps between the segments ensure that the soft magnetic iron elements match the diameter of the aluminium wheel at the 80K operating temperature.

As a new driving principle, a prototype of the FPW has been built as a quarter segment of the full wheel. The segment has been tested in the ambient and cryogenic environment to verify its functionality and performance, as described in the following section. 

   \begin{figure}[t!]
   \begin{center}
   \begin{tabular}{c} 
   \includegraphics[width=0.5\textwidth]{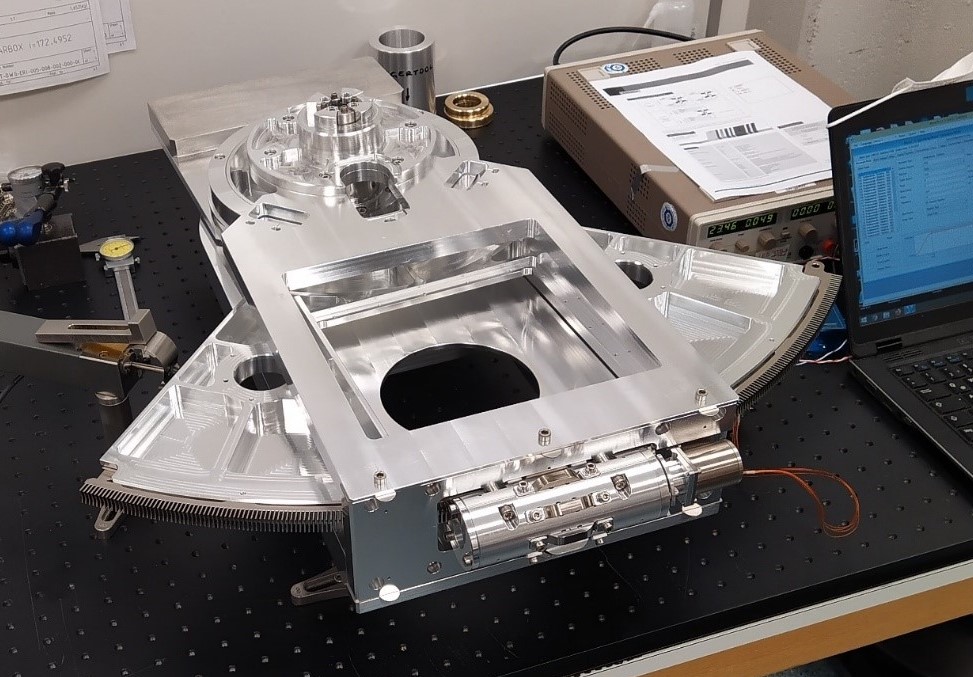}
   \end{tabular}
   \end{center}
   \caption{Photo of the FPW prototype segment.  \label{fig:prototype}}
   \end{figure} 

\section{The FPW prototype performance tests}
\label{tests}

To test the driving principle of the MCWG drive, a ptototype of the FPW was built (Fig.~\ref{fig:prototype}). The unit consists of a quarter wheel and a single worm drive. Building only a segment of the wheel allowed the parts to be machined in-house and the whole setup to be fitted more easily into the test cryostat. 

\subsection{Warm tests}

\begin{figure}[t!]
   \begin{center}
   \begin{tabular}{c} 
   \includegraphics[width=0.475\textwidth]{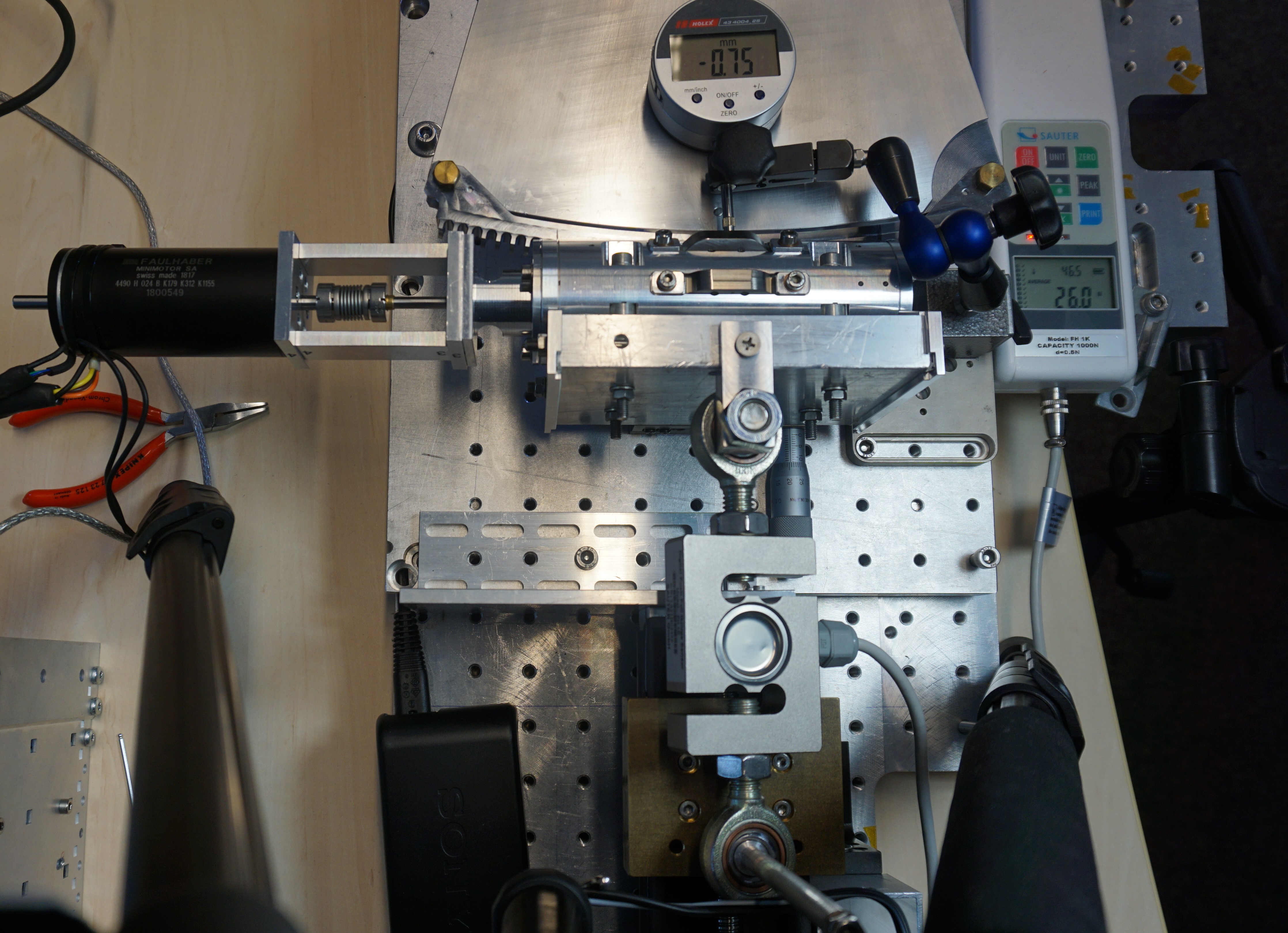}
   \includegraphics[width=0.44\textwidth]{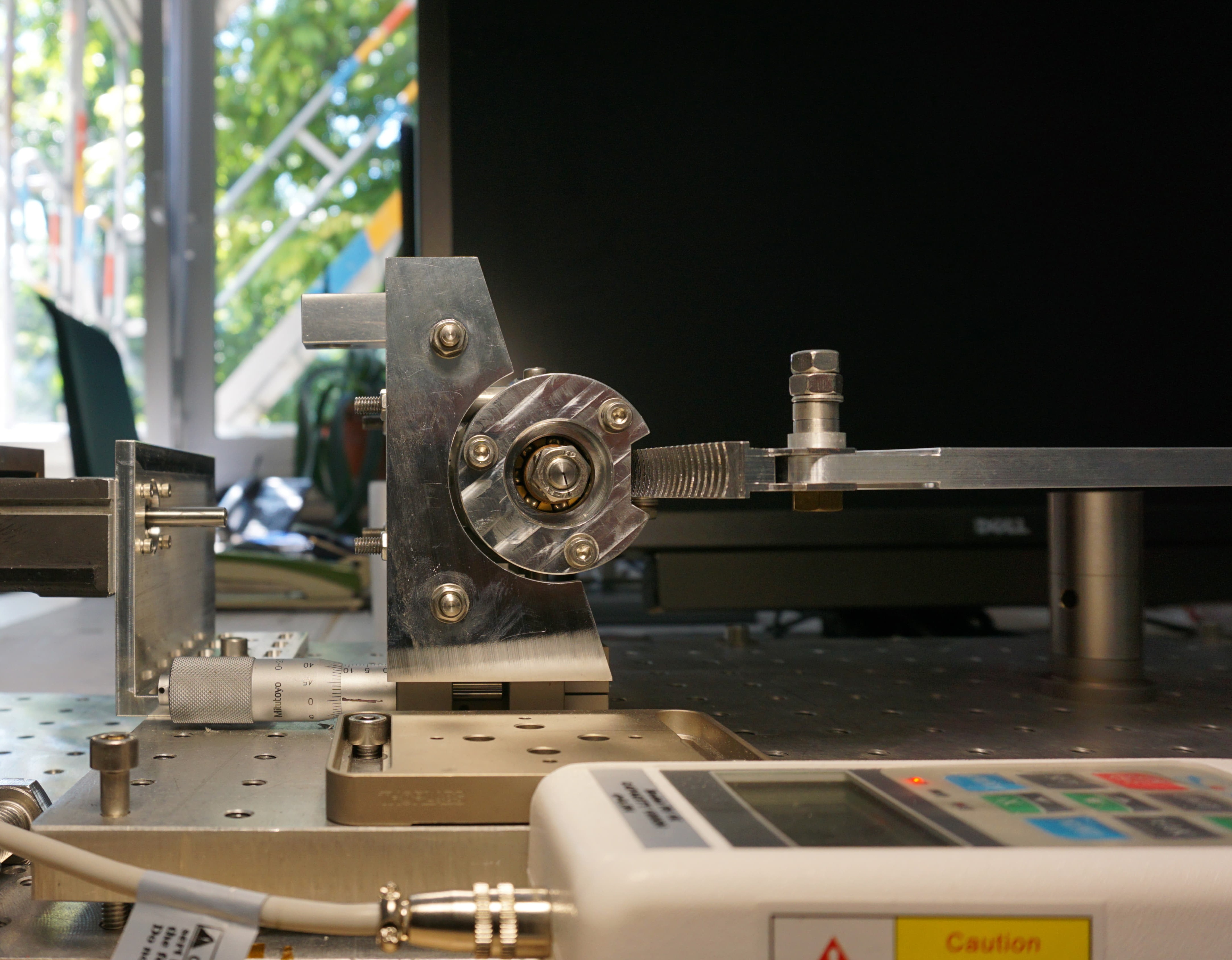}
   \end{tabular}
   \end{center}
   \caption{The top (left) and side (right) views of the force measurement setup described in Sect.~\ref{tests}. \label{fig:force_setup}}
   \end{figure}

\begin{figure}[t!]
   \begin{center}
   \begin{tabular}{c} 
   \includegraphics[width=0.47\textwidth]{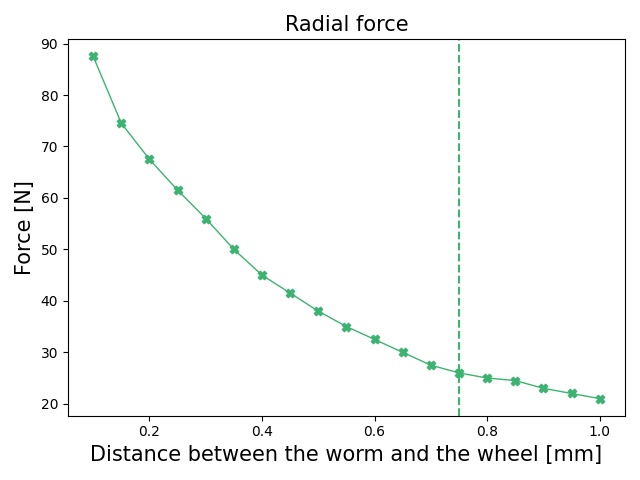}
   \includegraphics[width=0.47\textwidth]{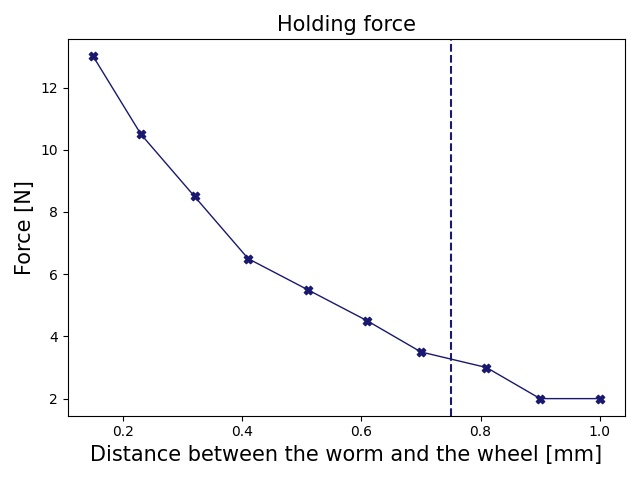}
   \end{tabular}
   \end{center}
   \caption{Measured radial (left) and holding (right) forces versus the distance between the worm and the wheel. Vertical lines indicate the gap of 0.75~mm selected for further testing. \label{fig:plot}}
   \end{figure}

The first step was to define an optimal gap between the worm and the wheel to prevent jamming in cryogenic conditions and excessive load on the bearings while still providing sufficient force to drive the mechanism. For this purpose, a test setup was built, as shown in Fig.~\ref{fig:force_setup}. In the setup, the worm unit is mounted on a linear micrometer stage allowing to vary the gap between the worm and the wheel. 
The distance feedback is provided by a portable coordinate measuring machine (CMM) arm. The setup comprises two digital force gauges: one to measure the radial force attracting the unit and another to measure the holding force. During the test, the gap was varied from 0.05~mm to 1~mm. 
The resulting measured radial and holding forces versus distance are shown in Fig.~\ref{fig:plot}. A gap of 0.75 mm was chosen for further prototype tests to minimize the load on the bearings, account for manufacturing tolerances, and ensure a "safe" gap to prevent the mechanism from getting stuck. However, it remains subject to final optimization on the full-wheel unit.  


\subsection{Cold tests}

The next step after the warm test was to verify the repeatability requirement for the FPW at its operational temperature of 80 K in a vacuum environment. The prototype mechanism was integrated into the PACS cryostat (a test cryostat available at MPE for different MICADO prototype and subsystem acceptance tests), which operates at 80 K. 

Since the prototype, as well as the final FPW and the AW units, are not equipped with cooling straps, the main cooling mechanism is radiation. Due to the relatively weak coupling of the unit to the main bench structure, an active thermal switch~\cite{ATS} (ATS) is used during cooldown. 
Prior the cooldown, the ATS is brought into contact with the wheel. 
After approximately 72~h of cooling, the prototype reaches its operating temperature of $\sim$80~K. Without the ATS, a cooldown would take around twice as long. When the ATS is in contact with the test unit, the temperature stability of the latter is about 0.003 K/h and a $\Delta$T from one side of the FPM to the other is about 0.07~K. During the wheel's operation, the contact between the wheel and ATS is interrupted to allow the wheel to move freely. The temperature stability, then, is about 0.116 K/h. 

   \begin{figure}[ht]
   \begin{center}
   \begin{tabular}{c} 
   \includegraphics[width=\textwidth]{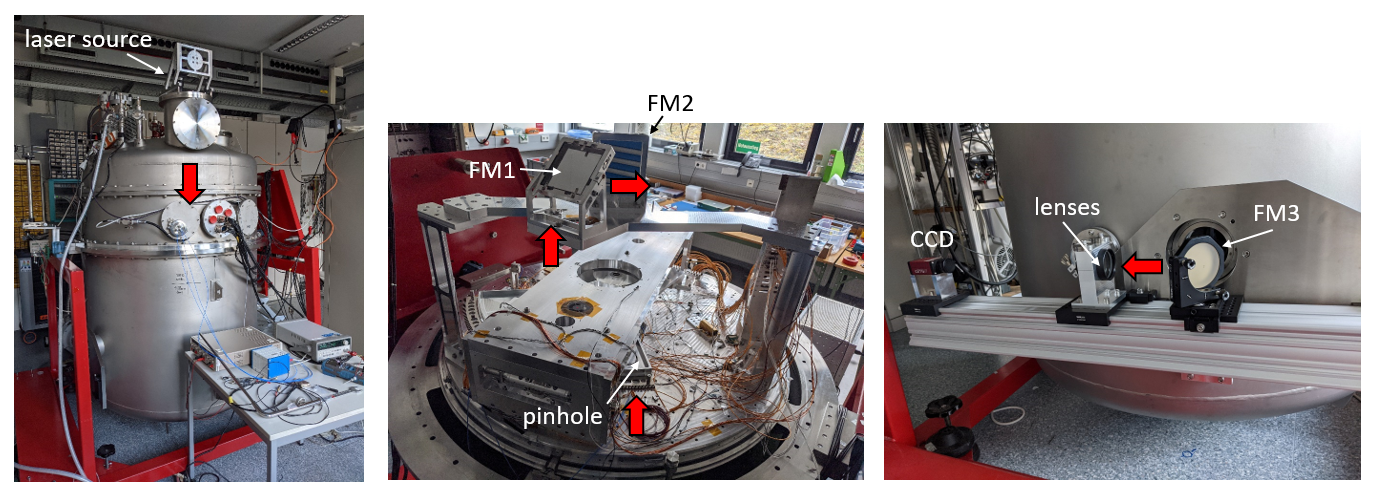}
   \end{tabular}
   \end{center}
   \caption{Optical setup for the FPW prototype testing. \textit{Left panel:} PACS cryostat in its operational orientation. The laser source is mounted on the top flange. \textit{Middle panel:} The prototype wheel is mounted to the cryostat bench. A small pinhole mounted to the wheel is hidden behind the mounting structure. Folding mirrors FM1 and FM2 are located above the wheel. \textit{Right panel:} The external optical setup consists of a folding mirror FM3, two lenses mounted in a tube, and a CCD camera. Red arrows illustrate the light path from the laser source to the CCD camera. \label{fig:optcal_setup}}
   \end{figure} 

The positioning requirement for the prototype wheel in the cryogenic environment was verified using two independent measurement approaches. First, the feedback from the capacitive distance sensors was used. The sensors have a few mm measurement range and 100 nm precision. Second, an external optical setup was built. It consists of a laser mounted to the top cryostat window, as shown in the left panel of Fig.~\ref{fig:optcal_setup}. The laser sends light inside the cryostat to illuminate a 20-micron pinhole mounted to the wheel. The light is then picked up by two folding mirrors (FM1 and FM2 in the middle panel of Fig.~\ref{fig:optcal_setup}) and is transferred through the second cryostat window outside, where it is picked by another flat mirror (FM3 in the right panel of Fig.~\ref{fig:optcal_setup}). This mirror then transfers the light through a system of two lenses, focusing it on a CCD camera (right panel of Fig.~\ref{fig:optcal_setup}), producing a diffraction-limited spot. As the wheel turns, the spot will move on the detector so that the position repeatability of the wheel can be measured. 

\begin{figure}[ht]
   \begin{center}
   \begin{tabular}{c} 
   \includegraphics[width=0.5\textwidth]{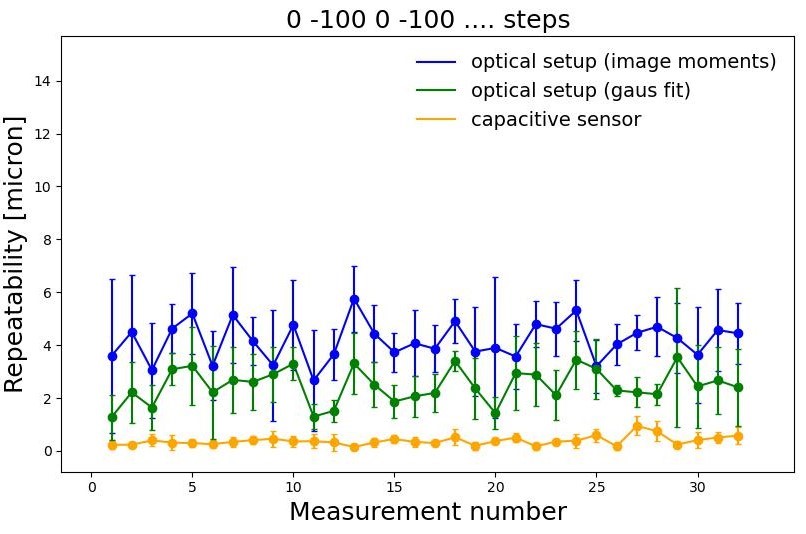}
   \includegraphics[width=0.5\textwidth]{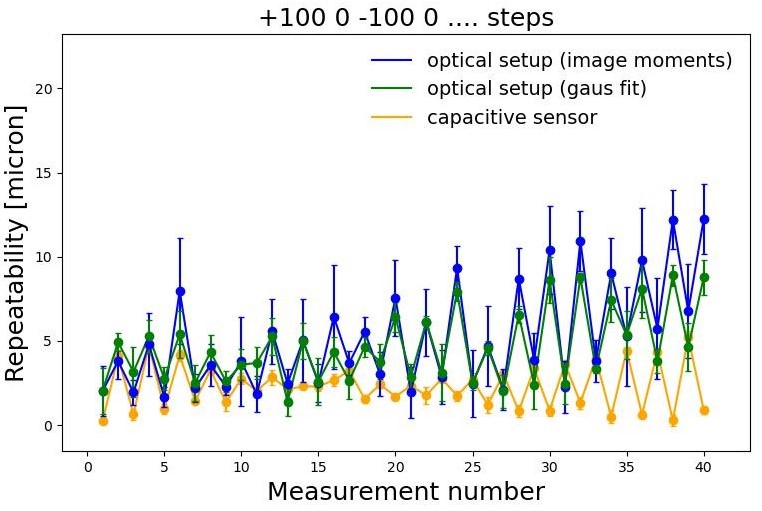}\\
   \includegraphics[width=0.5\textwidth]{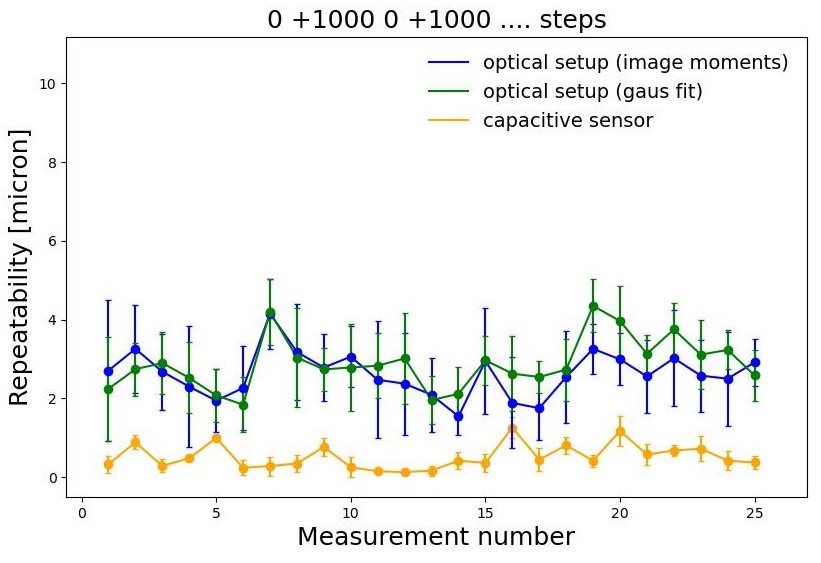}
   \includegraphics[width=0.5\textwidth]{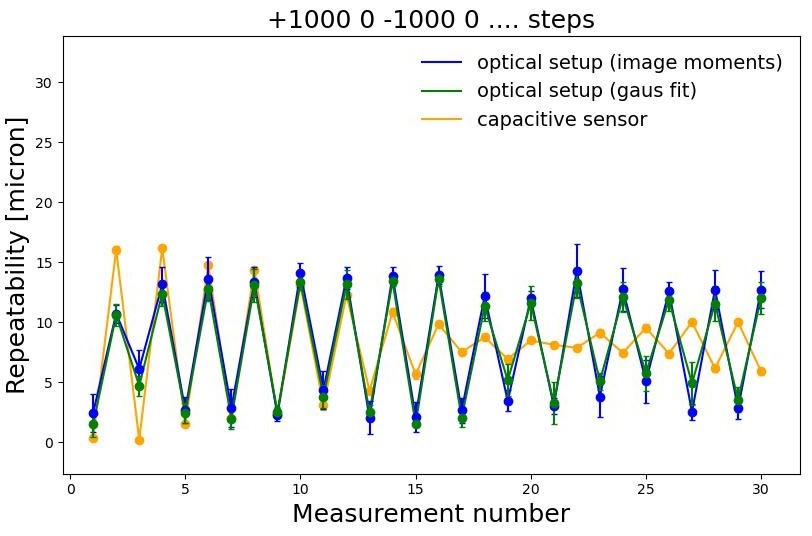}
   \end{tabular}
   \end{center}
   \caption{Examples of repeatability measurements of the FPW prototype: the position of the mechanism when moving 100 steps (top) and 1000 steps (bottom) in a single direction (left) or both directions (right). Orange illustrates capacitive sensor measurements. Blue and green illustrate optical measurements while applying calculations of the image moments or Gaussian fit, respectively, during image processing. \label{fig:test_results}}
   \end{figure}

The repeatability of the FPW prototype was verified by moving the wheel multiple times from a random initial position by 100 and 1000 steps in only positive, only negative, or both directions. Figure~\ref{fig:test_results} and Table~\ref{tab:repeatability} summarize the main results of the cold test campaign. On average, a 5$\mu$m RMS repeatability was achieved, with better performance when approaching a position always from the same side, due to some inherent hysteresis in the setup. The accuracy of the optical measurements was affected by vibrations present in the PACS cryostat from pumps and the laboratory environment. This explains the fact that capacitive sensor measurements are generally better than optical. Netherveless, Fig.~\ref{fig:test_results} and Table~\ref{tab:repeatability} clearly illustrate that FPW prototype fulfills its 15$\mu$m RMS specification with significant margin.

\section{Summary and conclusions}

This paper described the design and functionality of the FPM of MICADO. The core of the mechanism is its drive system based on magnetically coupled worm gear. This novel technology was implemented in a quarter-wheel prototype of the Focal Plane Wheel, optimized in warm conditions, and extensively tested in cryogenic conditions. It was shown that the repeatability requirement of 15$\mu$m RMS can easily be met. The manufacturing, assembly, and test of the full-size FPM is foreseen throughout 2024 - 2025. To conclude, the use of magnetically coupled worm gear drives is ideal for vacuum and cryogenic environments, providing contactless, wear-free operation and expanding the lifespan of moving mechanisms.

\newpage

   \begin{table}[ht!]
   \vspace{2.00mm}
\caption{Measured RMS repeatability (in $\mu$m) of the FPW prototype during cold tests.} 
\label{tab:repeatability}
\begin{center}       
\begin{tabular}{ p{0.2\linewidth} | p{0.2\linewidth} |p{0.2\linewidth} | p{0.2\linewidth}}
\hline 
\rule[-1ex]{0pt}{3.5ex}  Test configuration & Optical setup (image moments) & Optical setup (Gaussian fit) &  Capacitive sensors \\
\hline 
 \rule[-1ex]{0pt}{3.5ex} + 100 steps &	 2.6 &  3.2 &	1.4 \\
 \rule[-1ex]{0pt}{3.5ex} - 100 steps &	4.2 &  2.6 &	0.3 \\
 \rule[-1ex]{0pt}{3.5ex} + 100 0 -100 steps &  6.0 & 5.1 & 2.7 \\
 \rule[-1ex]{0pt}{3.5ex} + 1000 steps &	 2.7 & 3.0 &	0.6 \\
 \rule[-1ex]{0pt}{3.5ex} - 1000 steps &	 3.6 &  3.1 &	4.1 \\
\rule[-1ex]{0pt}{3.5ex} + 1000 0 -1000 steps &	 9.6 & 9.2 & 9.1 \\
\hline 
\end{tabular}
\end{center}
\end{table}

\newpage

\bibliography{report} 

\begin{thebibliography}{10}

\bibitem{MICADO}
{Sturm}, E., {Davies}, R., {Alves}, J., et~al., ``{The MICADO first light
  imager for the ELT: overview and current status},'' {\em Proc. SPIE} {\bf
  13096-37} (2024).

\bibitem{10.1117/12.2631613}
Tamai, R., Koehler, B., Cirasuolo, M., et~al., ``{Status of the ESO's ELT
  construction},'' {\em Proc. SPIE} {\bf 12182},  121821A (2022).

\bibitem{2020SPIE11445E..1GL}
{Larringan}, A., {M{\'u}gica}, A., {Murga}, G., et~al., ``{The final design for
  the Extremely Large Telescope prefocal stations},'' {\em Proc. SPIE} {\bf
  11445},  114451G (2020).

\bibitem{10.1117/12.2628969}
Ciliegi, P., Agapito, G., Aliverti, M., et~al., ``{MAORY/MORFEO at ELT: general
  overview up to the preliminary design and a look towards the final design},''
  {\em Proc. SPIE} {\bf 12185},  1218514 (2022).

\bibitem{SCAO}
{Clen\'et}, Y., {Gendron}, E., {Vidal}, F., et~al., ``{The MICADO first light
  imager for the ELT: first steps of the SCAO system MAIT},'' {\em Proc. SPIE}
  {\bf 13097-81} (2024).

\bibitem{cryostat}
{Emslander}, A., {Lang}, F., {Barl}, L., et~al., ``{The MICADO first light
  imager for the ELT: the mechanical manufacturing design of the cryostat},''
  {\em Proc. SPIE} {\bf 13094-123} (2024).

\bibitem{RO}
{Barboza}, S., {Bon\'e}, A., {Rohloff}, R.-R., et~al., ``{The MICADO first
  light imager for the ELT: starting manufacturing of the Relay Optics},'' {\em
  Proc. SPIE} {\bf 13096-204} (2024).

\bibitem{MCA}
{Setterholm}, B.~R., {Ramos}, J.~R., {M\"unch}, N., et~al., ``{The MICADO first
  light imager for the ELT: ongoing realization of the MICADO calibration
  assembly},'' {\em Proc. SPIE} {\bf 13096-207} (2024).

\bibitem{10.1117/12.2311867}
Schubert, J., Hartl, M., H{\"o}rmann, V., et~al., ``{The MICADO first light
  imager for ELT: cold optics instrument},'' {\em Proc. SPIE} {\bf 10702},
  107028W (2018).

\bibitem{MSM}
{Monna}, A., {Lang}, F., {Lange}, J., et~al., ``{The MICADO first light imager
  for the ELT: through the MAIT phase of the cryogenic Main Selection
  Mechanism},'' {\em Proc. SPIE} {\bf 13096-205} (2024).

\bibitem{2020SPIE11451E..3RR}
{Romp}, R., {Janssen}, A.~W., {Kunst}, P., et~al., ``{Design update of the
  central wheel mechanism},'' {\em Proc. SPIE} {\bf 11451},  114513R (2020).

\bibitem{detector}
{Bezawada}, N., {Haug}, M., {Klein}, B., et~al., ``{The MICADO first light
  imager for the ELT: Overview of the near infrared mosaic detector
  subsystem},'' {\em Proc. SPIE} {\bf 13103-24} (2024).

\bibitem{DPS}
{Soenmez}, A., {Kravchenko}, K., {Honsberg}, M., et~al., ``{The MICADO First
  Light Imager for the ELT: The Detector Positioning System of MICADO},'' {\em
  Proc. SPIE} {\bf 13100-115} (2024).

\bibitem{mcwg}
{Rabien}, S., ``{Magnetically coupled drive arrangement},'' {\em Patent
  application PCT/EP2021/059686} (2021).

\bibitem{ATS}
{Rabien}, S. and {Barl}, L., ``{Thermal Switching Device},'' {\em Patent
  application PCT/EP2021/059689} (2021).

\end{thebibliography}
\bibliographystyle{spiebib} 

\end{document}